\documentclass[11pt,a4paper]{article}
\usepackage[hmargin=1.0in]{geometry}
\usepackage{amsmath,amsfonts,graphicx,subfigure,color,multicol}
\usepackage{empheq}
\usepackage{amssymb}
\usepackage{epstopdf}
\usepackage{float,topcapt}
\usepackage{hyperref}
\usepackage{multirow}
\usepackage{rotating}
\usepackage{longtable}
\usepackage{algorithm}
\usepackage{algpseudocode}
\usepackage{relsize}
\usepackage{pdflscape}
\usepackage{placeins}
\usepackage[font=small,labelfont=bf]{caption}

\begin{document}



\newcommand{\nablav}{\vec{\nabla}}
\newcommand{\xv}{\vec{x}}
\newcommand{\wb}{\mathbf{w}}
\newcommand{\sbb}{\mathbf{s}}
\newcommand{\fbv}{\vec{\textbf{f}}}
\newcommand{\ximap}{\vec{\xi}}
\newcommand{\wbmap}{\wb^{\xiexp}}
\newcommand{\xiexp}{\hspace{0.04cm}\ximap}
\newcommand{\fbvmap}{\fbv^{\xiexp}}
\newcommand{\Jmap}{\vec{\vec{\left.\mathrm{J}\right.}}}
\newcommand{\nablavmap}{\vec{\nabla}^{\xiexp}}

\newcommand{\Wb}{\mathbf{W}}
\newcommand{\Fbv}{\vec{\textbf{F}}}
\newcommand{\Fbvmap}{\vec{\textbf{F}}^{\xiexp}}
\newcommand{\Fbvnum}{\vec{\textbf{F}}_\mathrm{num}^{\xiexp}}
\newcommand{\nmap}{\vec{n}^{\xiexp}}

\newcommand{\rhob}{{\rho}_{0}}
\newcommand{\pb}{{p}_{0}}
\newcommand{\cb}{{c}_{0}}

\newcommand{\Real}{\mathbb{R}}
\newcommand{\Complex}{\mathbb{C}}

\newcommand{\yb}{\mathbf{y}}
\newcommand{\eb}{\mathbf{e}}
\newcommand{\Ab}{\mathbf{A}}
\newcommand{\bb}{\mathbf{b}}

\newcommand{\ub}{\mathbf{u}}
\newcommand{\Fb}{\mathbf{F}}
\newcommand{\Lb}{\mathbf{L}}

\newcommand{\dt}{\Delta t}
\newcommand{\dtm}{\nu_{\mathrm{stab}}}
\newcommand{\dta}{\nu_{\mathrm{acc}}}
\newcommand{\dr}{\Delta r}

\newcommand{\abs}[1]{\vert #1 \vert}
\newcommand{\maxnorm}[1]{\vert \vert #1 \vert \vert_{\infty}}

\newcommand{\ndof}{N^\text{DOF}}
\newcommand{\Oop}{\mathcal O}

\newcommand{\chistab}{\chi_{\mathrm{stab}}}
\newcommand{\chiacc}{\chi_{\mathrm{acc}}}

\newcommand{\e}[1]{\ensuremath{\times 10^{#1}}}

\title{Large eddy simulation of a muffler with the high-order spectral difference method}
\author{Matteo Parsani\thanks{Current Institution: Computational Aerosciences Branch, NASA Langley Research Center, Hampton, VA 23681, USA (\mbox{matteo.parsani@nasa.gov})} \thanks{Division of Computer, Electrical and Mathematical Sciences \& Engineering, King Abdullah University of Science and Technology, Thuwal, 23955-6900, KSA} \and Michael Bilka\thanks{University of Notre Dame, Department of Aerospace and Mechanical Engineering, 365 Fitzpatrick Hall, Notre Dame, IN 46556-5637, USA (\mbox{michael.bilka.1@nd.edu})} \and Chris Lacor\thanks{Department of Mechanical Engineering, Vrije Universiteit Brussel, Pleinlaan 2, 1050 Brussels, Belgium (\mbox{chris.lacor@vub.ac.be})}}

\maketitle

\begin{abstract}The combination of the high-order accurate spectral difference discretization on 
unstructured grids with subgrid-scale modelling is investigated for large eddy 
simulation of a muffler at $Re = 4.64 \cdot 10^{4}$ and low Mach number. The 
subgrid-scale stress tensor is modelled by the wall-adapting local eddy-viscosity 
model with a cut-off length which is a decreasing function of the order of accuracy 
of the scheme. Numerical results indicate that although the
high-order solver without subgrid-scale modelling is already able to capture well the 
features of the flow, the coupling with the wall-adapting local eddy-viscosity model improves the quality 
of the solution.
\end{abstract}

\section{Introduction}
\label{sec:Intro}
Throughout the past two decades, the development of high-order accurate spatial 
discretization has been one of the major fields of research in computational
fluid dynamics (CFD), computational aeroacoustics (CAA), computational 
electromagnetism (CEM) and in general computational physics
characterized by linear and nonlinear wave propagation phenomena. 
High-order accurate discretizations have the potential to improve the computational 
efficiency required to achieve a desired error level. In fact, compared with low order schemes, high order methods offer better wave propagation properties and increased accuracy for a comparable number of degrees of freedom (DOFs). Therefore, it may be advantageous to use such schemes for problems that require very low numerical dissipation and small error levels \cite{Wang-book-2011}. Moreover, since computational science is increasingly used as an industrial
design and analysis tool, high accuracy must be achieved on unstructured 
grids which are required for efficient meshing. These needs have been the driving
force for the development of a variety of higher order schemes for unstructured meshes such as 
the Discontinuous Galerkin (DG) method \cite{busch-DG-Laser-2011,hesthaven-nodalDG-2007}, the Spectral 
Volume (SV) method \cite{wang-SV-2006}, the Spectral Difference (SD) method 
\cite{may-SD-2006,sun-SD-2007}, the Energy Stable Flux Reconstruction 
\cite{castonguay-FR-2011} and many others.

In this study we focus on a SD solver for unstructured hexahedral grids
(tensorial cells). The SD method has been proposed as an alternative high order
collocation-based method using local interpolation of the strong form of the 
equations. Therefore, the SD scheme  has an important advantage over classical DG 
and SV methods, that no integrals have to be evaluated to compute the 
residuals, thus avoiding the need for costly high-order accurate quadrature 
formulas.
    
Although the formulation of high-order accurate spatial discretization is now fairly 
mature, their application for the simulation of general turbulent flows implies
that particular attention has still to be paid to subgrid-scale (SGS) models. 
So far, the combination of the SD method with SGS models for LES has not been 
widely investigated. In 2010, Parsani et al. \cite{parsani-2DSD-LES-2010} reported
the first implementation in study of a two-dimensional (2D) third-order accurate SD solver 
coupled with the Wall-Adapting Local Eddy-viscosity (WALE) model \cite{nicoud-WALE-1999}
and a cut-off length which is a decreasing function of the order of accuracy. 
A successful extension of that approach to a three-dimensional (3D) second-order accurate SD solver
has been reported in \cite{parsani-3DSD-LES-2011}.
Very recently, Lodato and Jameson \cite{lodato-3DSD-LES-2012} have presented 
an alternative technique to model the unresolved scales in the flow field: A structural 
SGS approach with the WALE Similarity Mixed model (WSM), where constrained explicit filtering represents 
a key element to approximate subgrid-scale interactions. The performance of 
such an algorithm has been also satisfactory.

In this study, we couple for the first time the approach proposed in \cite{parsani-2DSD-LES-2010}
with a 3D fourth-order accurate SD solver, for the simulation of the turbulent 
flow in an industrial-type muffler at $Re = 4.64 \cdot 10^{4}$.
The goal is to investigate if the coupling of a high-order SD scheme with a sub-grid 
closure model improves the quality of the results when the grid resolution is 
relatively low. The latter requirement is often desirable when a high-order accurate
spatial discretization is used.

\section{Physical model and numerical algorithm}
\label{sec:num-algo}
In this study the system of the Navier-Stokes equations for a compressible 
flow are discretized in space using the SD method and the subgrid-scale stress tensor is modelled by the WALE approach.

\subsection{Filtered Navier-Stokes equations}
\label{subsec:LES-equations}
The three physical conservation
laws for a general Newtonian fluid, i.e., the continuity, the
momentum and energy equations, are introduced using the following
notation: $\rho$ for the mass density, $ \vec{u}\in
\mathbb{R}^{dim}$ for the velocity vector in a physical space with
$dim$ dimensions, $P$ for the static pressure and $E$ for the
specific total energy which is related to the pressure and the
velocity vector field by $E = \frac{1}{\gamma - 1}\frac{P}{\rho} + \frac{\vert \vec{u}\vert^2}{2}$,
where $\gamma$ is the constant ratio of specific heats and it is $1.4$ for air in standard conditions.

The system, written in divergence form and
equipped with suitable initial-boundary conditions, is
\begin{equation}
\frac{\partial \textbf{w}}{\partial t} + \vec{\nabla}\cdot
\left(\vec{\textbf{f}}_C\left(\textbf{w}\right) -
\vec{\textbf{f}}_D\left(\textbf{w},\vec{\nabla}
\textbf{w}\right)\right) = \frac{\partial \textbf{w}}{\partial t} +
\vec{\nabla}\cdot \vec{\textbf{f}} = 0, \label{eq:conseq}
\end{equation}
where $\textbf{w} = \left(\overline{\rho}, \overline{\rho}
\tilde{\vec{u}}, \overline{\rho} \tilde{E}\right)^T$ is the vector of the filtered
conservative variables and $\vec{\textbf{f}}_C = \vec{\textbf{f}}_C\left(\textbf{w}\right)$ and
$\vec{\textbf{f}}_D = \vec{\textbf{f}}_D\left(\textbf{w},\vec{\nabla} \textbf{w}\right)$ represent
the convective and the diffusive fluxes, respectively. Here the symbols 
$(\overline{\cdot})$ and $(\tilde{\cdot})$
represent the spatially filtered field and the Favre
filtered field defined as $\tilde{\vec{u}} = \overline{\rho
\vec{u}}/\overline{\rho}$.

In a general
3D ($dim =3$) Cartesian space, $\vec{x} = \left[x_1, x_2,
x_3\right]^T$, the components of the flux vector $\vec{\textbf{f}}\left(\textbf{w},\vec{\nabla} \textbf{w}\right) =
\left[\textbf{f}_1, \textbf{f}_2, \textbf{f}_3\right]^T$ are given by
\begin{equation*}
  \textbf{f}_1 = \left(\begin{array}{c}\overline{\rho} \tilde{u}_1 \\ \overline{\rho} \tilde{u}_1^2+\overline{P} - \tilde{\sigma}_{11} + \tau_{11}^{sgs}  \\
  \overline{\rho} \tilde{u}_1 \tilde{u}_2 - \tilde{\sigma}_{21} + \tau_{21}^{sgs} \\ \overline{\rho} \tilde{u}_1 \tilde{u}_3 - \tilde{\sigma}_{31} + \tau_{31}^{sgs} \\
  \tilde{u}_1 \left(\overline{\rho}\tilde{E}+\overline{P}\right) - \tilde{u}_1 \left(\tilde{\sigma}_{11} - \tau_{11}^{sgs}\right) - \tilde{u}_2\left(\tilde{\sigma}_{21} - \tau_{21}^{sgs}\right)-\tilde{u}_3\left(\tilde{\sigma}_{31} - \tau_{31}^{sgs} \right)-c_P \frac{\mu}{Pr}\frac{\partial \tilde{T}}{\partial x_1} + q_{1}^{sgs}\end{array}\right),
\end{equation*}
\begin{equation*}
\textbf{f}_2 = \left(\begin{array}{c}\overline{\rho} \tilde{u}_2 \\ \overline{\rho} \tilde{u}_1 \tilde{u}_2 - \tilde{\sigma}_{12}
+ \tau_{12}^{sgs}\\ \overline{\rho} \tilde{u}_2^2+\overline{P} - \tilde{\sigma}_{22} + \tau_{22}^{sgs}\\
\overline{\rho} \tilde{u}_2 \tilde{u}_3 - \tilde{\sigma}_{32} + \tau_{32}^{sgs} \\ 
\tilde{u}_2 \left(\overline{\rho}\tilde{E}+\overline{P}\right) - \tilde{u}_1 \left(\tilde{\sigma}_{12} - \tau_{12}^{sgs} \right) -  \tilde{u}_2 \left(\tilde{\sigma}_{22} - \tau_{22}^{sgs} \right) - \tilde{u}_3 \left(\tilde{\sigma}_{32} - \tau_{32}^{sgs}\right) 
 - c_P \frac{\mu}{Pr}\frac{\partial \tilde{T}}{\partial x_2} + q_{2}^{sgs}\end{array}\right),
\end{equation*}
\begin{equation*}
\textbf{f}_3 = \left(\begin{array}{c}\overline{\rho} \tilde{u}_3 \\ \overline{\rho} \tilde{u}_1 \tilde{u}_3 - \tilde{\sigma}_{13} + \tau_{13}^{sgs}\\
\overline{\rho} \tilde{u}_2 \tilde{u}_3 - \tilde{\sigma}_{23} + \tau_{23}^{sgs} \\ \overline{\rho} \tilde{u}_3^2 + \overline{P} - \tilde{\sigma}_{33} + \tau_{33}^{sgs} \\
\tilde{u}_3 \left(\overline{\rho}\tilde{E}+\overline{P}\right) - \tilde{u}_1 \left(\tilde{\sigma}_{13} - \tau_{13}^{sgs}\right) - \tilde{u}_2 \left(\tilde{\sigma}_{23} - \tau_{23}^{sgs}\right) - \tilde{u}_3 \left(\tilde{\sigma}_{33} - \tau_{33}^{sgs}\right)
- c_P \frac{\mu}{Pr}\frac{\partial \tilde{T}}{\partial x_3} + q_{3}^{sgs}\end{array}\right),
\end{equation*}
where $c_P$, $\mu$, $Pr$ and $T$ represent respectively the specific heat capacity at constant pressure,
the dynamic viscosity, the Prandtl number and the temperature of the fluid. Moreover, $\sigma_{ij}$
represents the $ij-$component of the resolved viscous stress tensor \cite{pope-Turbulence}. 


Both momentum
and energy equations differ from the classical fluid dynamic equations only for two
terms which take into account the contributions from the unresolved scales. These contributions,
represented by the specific subgrid-scale stress tensor $\tau_{ij}^{sgs}$ and by the subgrid heat flux vector defined $q_{i}^{sgs}$, appear when the spatial filter is applied to the
convective terms \cite{pope-Turbulence}.
The interactions of $\tau_{ij}^{sgs}$ and $q_{i}^{sgs}$ with the resolved scales have to be
modeled through a subgrid-scale closure model because they cannot be determined using only the
resolved flow field $\textbf{w}$.

\subsubsection{The wall-adapted local eddy-viscosity closure model}
\label{subsubsec:WALE}
The smallest scales present in the flow field and their
interaction with the resolved scales have to be modeled through the
subgrid-scale term $\tau_{ij}^{sgs}$. The most common
approach to model such a tensor is based on the eddy-viscosity concept 
in which one assumes that the residual stress is proportional to a measure of 
the filtered local strain rate \cite{pope-Turbulence}, which is defined as follows:
\begin{equation}\label{eq:Eddy-ViscSGSMomentum1}
\tau_{ij}^{sgs} - \tau_{kk}^{sgs} \delta_{ij} = -2 \, \overline{\rho} \,
\nu_t \left(\tilde S_{ij} - \frac{\delta_{ij}}{3} \tilde S_{kk}\right).
\end{equation}

In the WALE model, it is assumed that the
eddy-viscosity $\nu_t$ is proportional to the square of the length
scale of the cut-off length (or width of the grid filter) and the filtered local rate of strain.
Although
the model was originally developed for incompressible flows, it
can also be used for variable density flows by giving the
formulation as follows
\begin{equation}\label{eq:WALESGSMomentum}
\nu_t = \left(C\Delta\right)^{2}\left\vert \tilde S \right\vert .
\end{equation}
Here $\left \vert \tilde S \right\vert$ is defined as
\begin{equation}\label{eq:Eddy-ViscSGSMomentum3}
\left\vert \tilde S \right\vert = \frac{\left[\tilde S_{ij}^d \,
\tilde S_{ij}^d\right]^{3/2}}{\left[\tilde S_{ij} \, \tilde S_{ij}\right]^{5/2}
+ \left[\tilde S_{ij}^d \, \tilde S_{ij}^d\right]^{5/4}},
\end{equation}
where $\tilde S_{ij}^d$ is the traceless symmetric part of the square of the 
resolved velocity gradient tensor $\tilde{g}_{ij} = \frac{\partial \tilde{u}_i}{\partial x_j}$.
Note that in Equation \eqref{eq:WALESGSMomentum} $\Delta$, i.e., the cut-off length, is an unknown function. Often the cut-off length is 
taken proportional to the smallest resolvable length scale
of the discretization. In the present work, the definition of the
grid filter function is given in Section \ref{subsec:SD}, where the SD method is 
discussed.

\subsection{Spectral difference method}
\label{subsec:SD}
Consider a problem governed by a general system of conservation laws
given by Equation \eqref{eq:conseq} and valid on a domain $\Omega \subset\mathbb{R}^{dim}$ 
with boundary 
$\partial\Omega$ and completed with consistent initial and boundary conditions. 
The domain is divided into $N$
non-overlapping cells, with cell index $i$.

In order to achieve an efficient implementation of the SD method, all hexahedral
cells in the physical domain are mapped into cubic elements using local 
coordinates $\vec{\xi} = \left[\xi_{1},\xi_{2},\xi_{3}\right]^T$. Such a 
transformation is characterized by the Jacobian matrix $\Jmap_i$ with determinant
$det(\Jmap_i)$. Therefore, system \eqref{eq:conseq} can be written in 
the mapped coordinate system as
\begin{equation}
\frac{\partial\wb_i^{\xiexp}}{\partial t} 
= -\frac{\partial\mathbf{f}_{1,i}^{\xiexp}}{\partial\xi_{1}}
-\frac{\partial\mathbf{f}_{2,i}^{\xiexp}}{\partial\xi_{2}}
-\frac{\partial\mathbf{f}_{3,i}^{\xiexp}}{\partial\xi_{3}} 
= -\nablavmap\cdot\fbvmap_i,
\label{eq:genHypSysMap}
\end{equation}
where $\wbmap_i \equiv det(\Jmap_i) \, \wb$ and $\nablavmap$ are the conserved variables and
the generalized divergence differential operator in the mapped coordinate system, respectively.

For a $\left(p+1\right)$-th-order accurate $dim$-dimensional scheme, 
$N^s$ \emph{solution collocation points} with index $j$ are introduced at positions 
$\ximap_j^s$ in each cell $i$, with $N^s$ given by $N^s=\left(p+1\right)^{dim}$.
Given the values at these points, a polynomial approximation of degree $p$ of
the solution in cell $i$ can be constructed. This polynomial is called the
\emph{solution polynomial} and is usually composed of a set of Lagrangian 
basis polynomial $L_j^s\left(\ximap\right)$ of degree $p$:
\begin{equation}
	    \Wb_i\left(\ximap\right) = \sum_{j=1}^{N^s} \Wb_{i,j} \, L_j^s\left(\ximap\right).
\label{eq:sdsolpoly}
\end{equation}
Therefore, the unknowns of the SD method are the interpolation coefficients $\Wb_{i,j} = \Wb_i\left(\ximap_j^s\right)$ which are the approximated values
of the conserved variables $\wb_i$ at the solution points.  

The divergence of the mapped fluxes 
$\nablavmap\cdot\fbvmap$ at the solution points is computed by 
introducing a set of $N^f$ flux collocation points with index $l$ and at positions 
$\ximap_l^f$, supporting a polynomial of degree $p+1$. The evolution of the 
mapped flux vector $\fbvmap$ in cell $i$ 
is then approximated by a flux polynomial $\Fbvmap_i$, which is obtained
by reconstructing the solution variables at the flux points and evaluating the 
fluxes $\Fbvmap_{i,l}$ at these points. The flux is also represented by a
Lagrange polynomial:
\begin{equation}
        \Fbvmap_i\left(\ximap\right) = \sum_{l=1}^{N^f} \Fbvmap_{i,l} \, L_l^f\left(\ximap\right),
\label{eq:sdfluxpoly}
\end{equation}
where the coefficients of the flux interpolation are defined as
\begin{equation}
    \Fbvmap_{i,l} =
        \begin{cases}
            \Fbvmap_i\left(\ximap_l^f\right), & \quad \ximap_l^f \in \Omega_i, \\
            \Fbvnum\left(\ximap_l^f\right), & \quad \ximap_l^f \in \partial\Omega_i.
        \end{cases}
\end{equation}
Here $\Fbvnum$ is the numerical flux vector at the cell interface. In fact, the 
solution at a face is in general not continuous and requires the solution of a Riemann 
problem to maintain 
conservation at a cell level (i.e., the flux component normal to 
a face $\Fbvnum \cdot \nmap$ must be continuous between two neighboring 
cells). Approximate Riemann solvers are typically used (e.g. Rusanov Riemann solver).
The tangential component of $\Fbvnum$ is usually taken from the interior cell.

Taking the divergence of the flux polynomial $\nablavmap\cdot\Fbvmap_i$ in the 
solution points results in the following modified form of 
\eqref{eq:genHypSysMap}, describing the evolution of the conservative
variables at the solution points:
\begin{equation}
\frac{d\Wb_{i,j}}{dt} = 
-\left.\nablav\cdot\Fbv_i\right|_j = -\frac{1}{J_{i,j}}\left.\nablavmap\cdot
\Fbvmap_i\right|_j = \mathbf{R}_{i,j},
\label{eq:residualsSD}
\end{equation}
where $\Fbv_i$ is the flux polynomial vector in the physical space whereas 
$\mathbf{R}_{i,j}$ is the SD residual associated with $\Wb_{i,j}$. This is
a system of ODEs, in time, for the unknowns $\Wb_{i,j}$. In this work, the
optimized explicit eighteen-stages fourth-order Runge-Kutta schemes presented in
\cite{parsani-optimalERK-2012} is used to solve such a system at each time step.

\subsubsection{Solution and flux points distributions}
In 2007, Huynh \cite{THuynh2007} showed
that for quadrilateral and hexahedral cells, tensor product flux point 
distributions based on a one-dimensional flux point distribution consisting of the end points
and the Legendre-Gauss quadrature points lead to stable schemes for arbitrary order of accuracy. In 2008, Van den Abeele et al.~\cite{Abeele2008a} showed an interesting
property of the SD method, namely that it is independent of the positions 
of its solution points in most general circumstances. This property implies 
an important improvement in efficiency, since the solution points can be placed 
at flux point positions and thus a significant number of solution reconstructions can be 
avoided. Recently, this property has been proved by Jameson 
\cite{Jameson2010}. 

\subsubsection{Cut-off length $\Delta$}
In Section \ref{subsubsec:WALE} we have seen
that in the WALE model the cut-off length $\Delta$ is used to
compute the turbulent eddy-viscosity $\nu_t$, i.e., Equation
(\ref{eq:WALESGSMomentum}). Following the approach presented in \cite{parsani-2DSD-LES-2010}, for each cell 
with index $i$
and each flux points with index $l$ and positions
$\boldsymbol{\xi}_l^f$, we use the following definition of
filter width
\begin{equation}
\Delta_{i,l} = \left[\frac{1}{N^s}det\left(\left.\vec{\vec{\left.\mathrm{J}\right.}}_i\right|_{\boldsymbol{\xi}_l^f}\right)\right]^{1/dim}
= \left(\frac{det(J_{i,l})}{N^s}\right)^{1/dim}.
\label{eq:filterWidthInternalFluxPoints}
\end{equation}
Notice that the cell filter width is not
constant in one cell, but it varies because the Jacobian matrix is a
function of the positions of the flux points. Moreover, for a given
mesh, the number of solution points depends on the order of the SD
scheme, so that the grid filter width decreases by increasing the polynomial 
order of the approximation.

\section{Numerical results}
\label{sec:results}
\begin{figure}[htbp!]
\centering
\scalebox{0.40}{\input{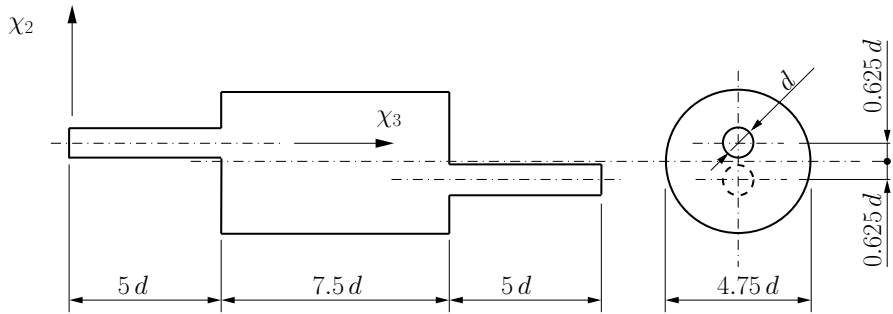}}
\caption{Configuration of the 3D muffler test case.}
\label{fig: 3D-VKI-Muffler-geo}
\end{figure}

The main purpose of this section is to evaluate the accuracy and the reliability
of the fourth-order SD-LES solver for simulating a 3D turbulent flow in an industrial-type muffler. The results are compared with the particle
image velocimetry (PIV) measurement performed at
the Department of Environmental and Applied Fluid Dynamics of the von Karman 
Institute for Fluid Dynamics \cite{bilka-PIV}.
In Figure \ref{fig: 3D-VKI-Muffler-geo}, the geometry
of the muffler and its characteristic dimensions are illustrated, where the
flow is from left to right.

At the inlet, mass density and velocity profiles are imposed. The inlet velocity profile in the $x_3$ direction is given by
\begin{equation*}
u_{3} = u_{max} \left\{\frac{1}{2} - \frac{1}{2} \tanh{ \left[2.2\left(\frac{r}{d/2}-\frac{d/2}{r}\right)\right]}\right\}.
\end{equation*}
At the outlet only the pressure is prescribed. In accordance to the experiments, 
the inlet Mach number and the Reynolds number, based on maximum velocity at the 
inlet $u_{max}$ and the diameter of the inlet/outlet $d$ ($d = 4 \, cm$), are set respectively 
to $M_{inlet}=0.05$ and $Re = 4.64 \cdot 10^{4}$. 
\\
\\
The flow is computed using fourth-order ($p=3$) SD scheme on a grid with $36,612$ 
hexahedral elements which was generated with the open source software Gmsh 
\cite{geuzaine-Gmsh-2009}. Second-order boundary elements are used to approximate
the curved geometry. The total number of DOFs is approximately 
$2.3 \cdot 10^{6}$ (i.e., $36,612 \cdot (p+1)^3$). The maximum CFL number used for the computations started 
from $0.1$ and increased up to $0.65$. After the flow field was fully developed, 
time averaging is performed for a period corresponding to about 25 flow-through 
times.

The computation is validated on the center plane of the expansion coinciding 
with the center planes of the inlet and outlet pipes using the PIV results from 
\cite{bilka-PIV}. All of the measurements 
are taken on the symmetrical center plane of the muffler. The reference cross section corresponds to the entrance of the expansion chamber. It should be noted that
the circular nature of the geometry acts as a lens causing a change in 
magnification in the radial direction ($x_2$) which prevents from capturing images close to the wall. It is found 
that outside $~1 \, cm$ from the wall the magnification effect is negligible and as 
the mean stream-wise direction is in the direction of constant magnification and 
has only little effect on the particle correlations no corrections are deemed necessary.

\begin{figure}[t!]
\centering
    \subfigure[\small{$1 d$ downstream}.]{\includegraphics[width = 0.490\columnwidth]{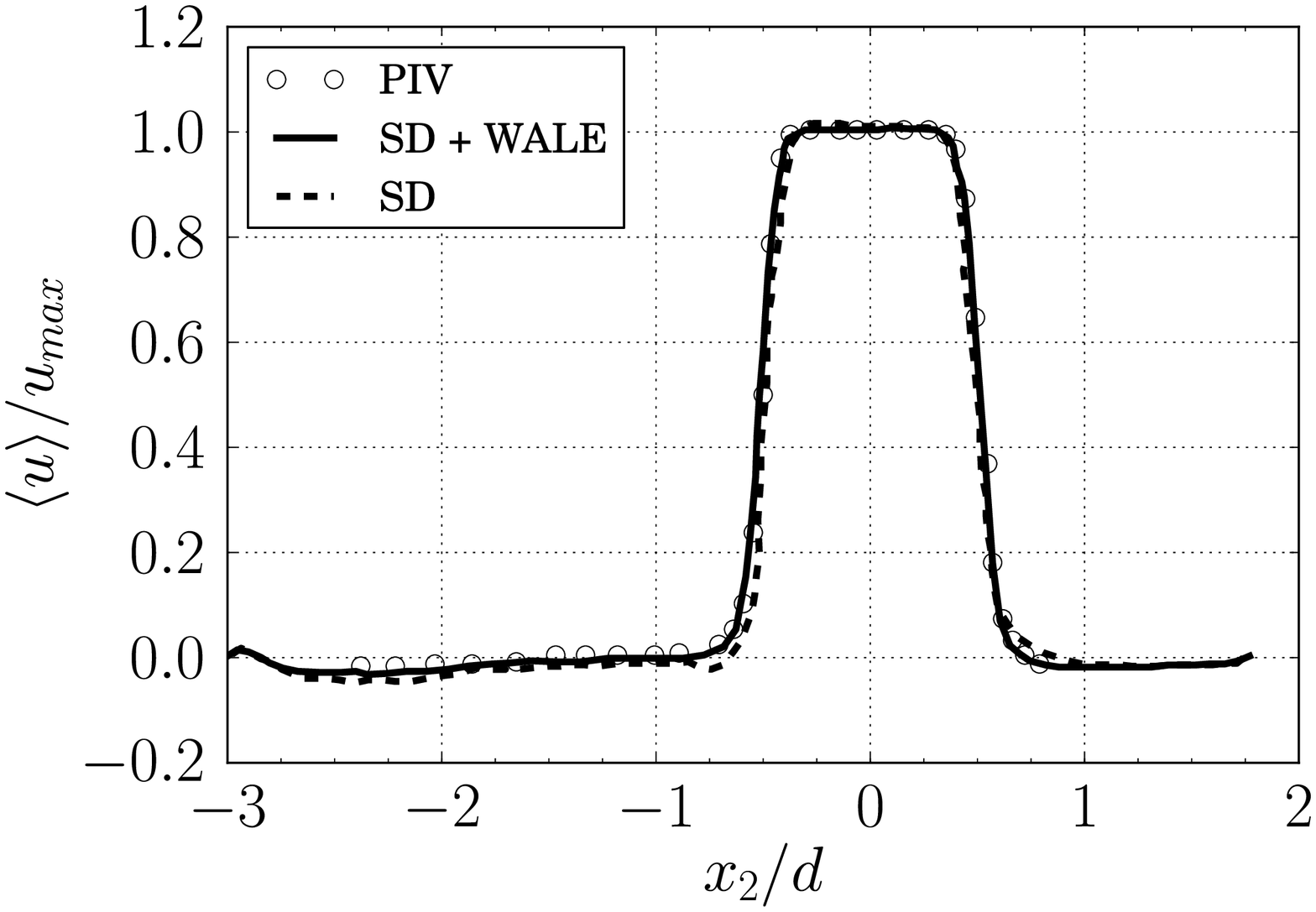}}  
	  \subfigure[\small{$4 d$ downstream}.]{\includegraphics[width = 0.490\columnwidth]{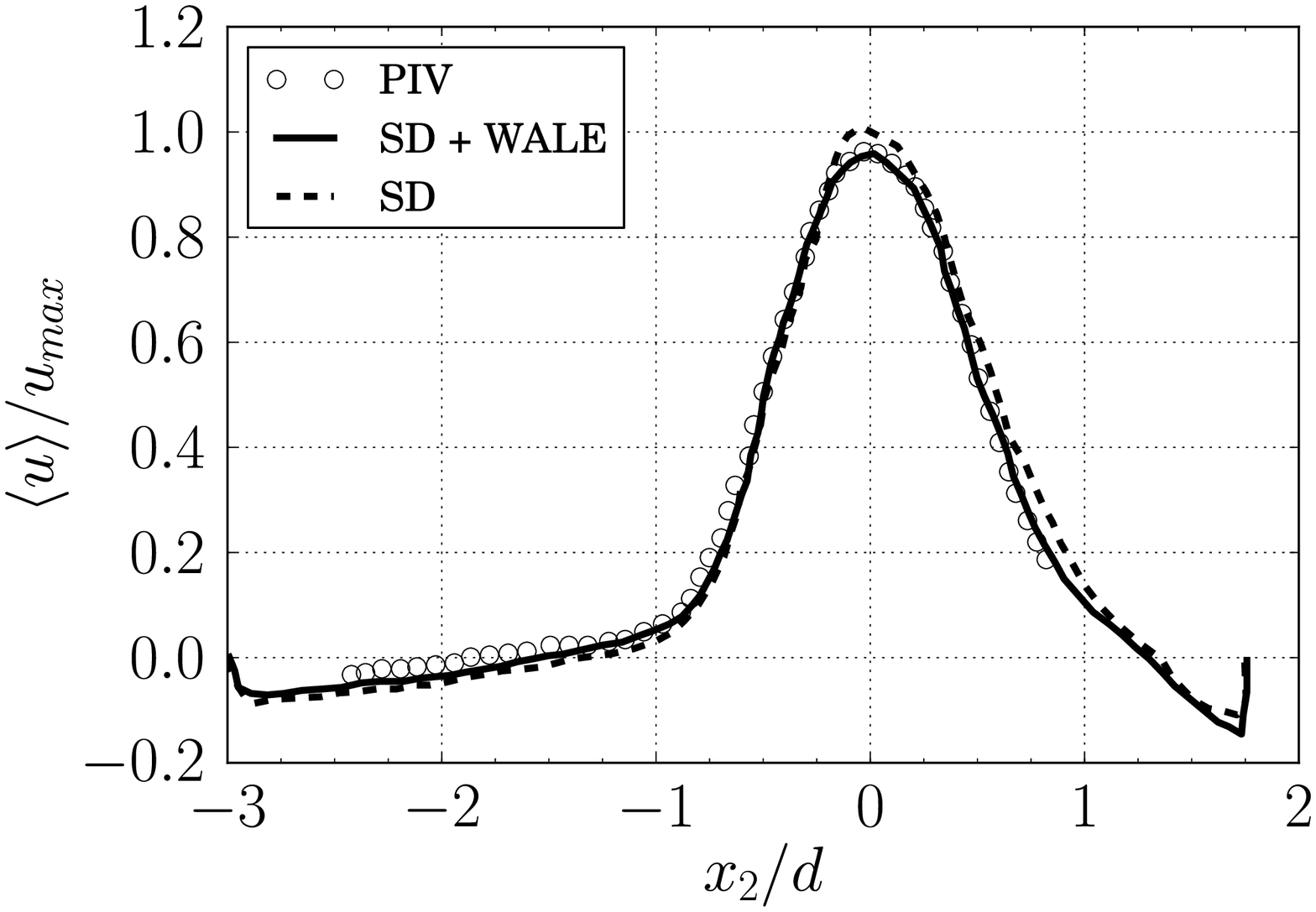}}
    \subfigure[\small{$6 d$ downstream}.]{\includegraphics[width = 0.490\columnwidth]{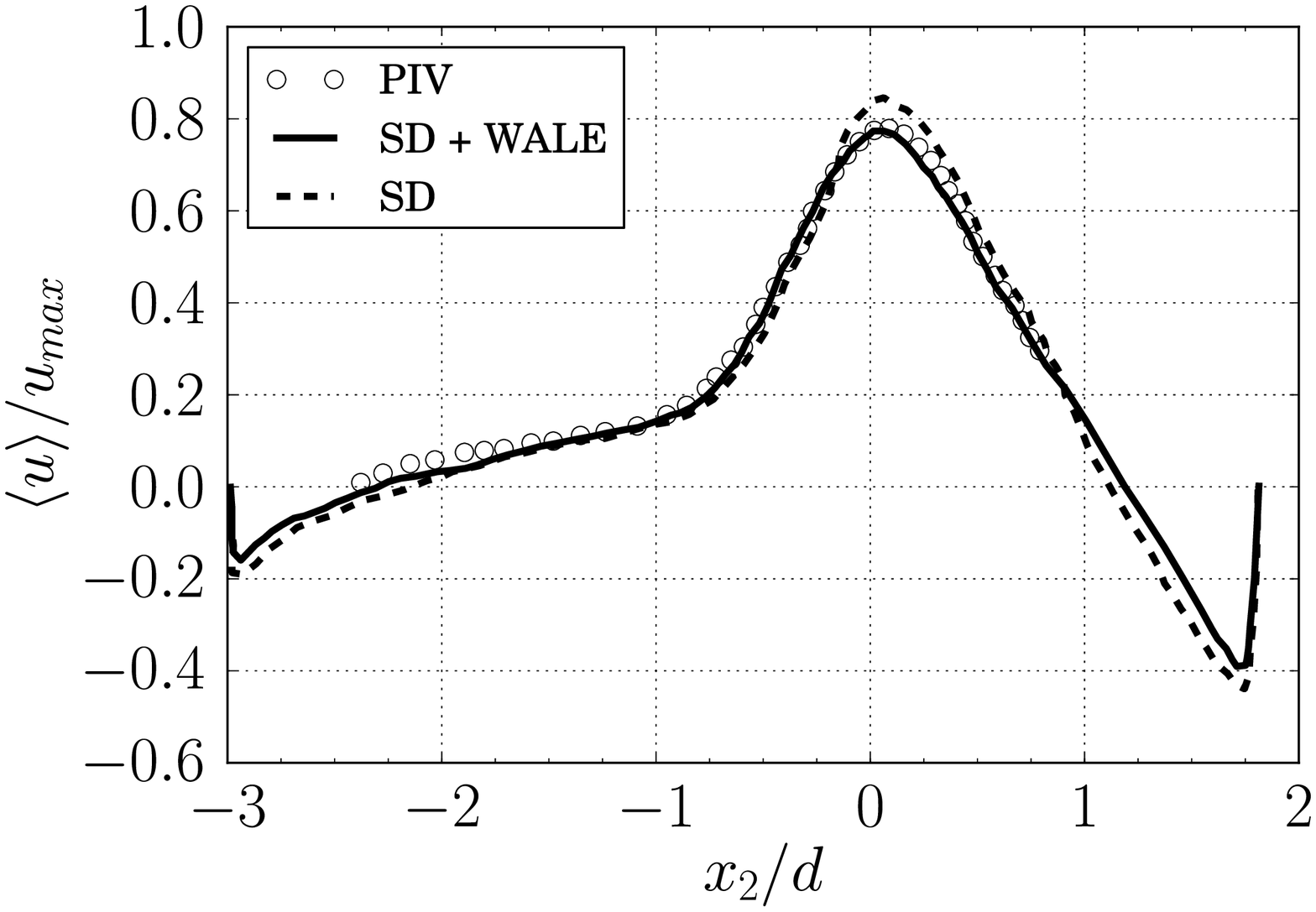}}  
    \subfigure[\small{$7 d$ downstream}.]{\includegraphics[width = 0.490\columnwidth]{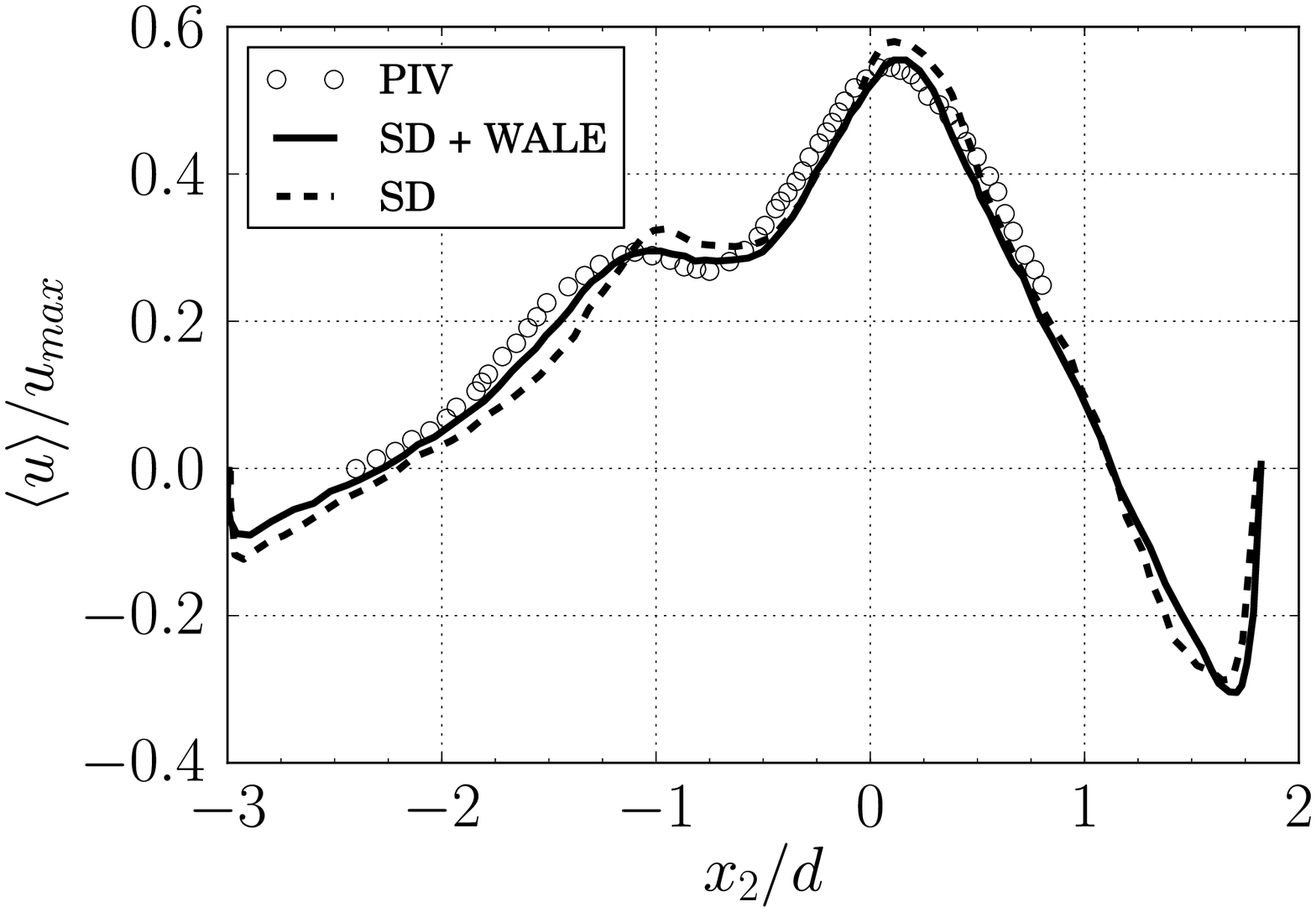}}
\caption{Time-averaged velocity profile in the axial direction $\langle\tilde{u}_3\rangle/u_{max}$ at four cross sections in the expansion chamber, obtained with fourth-order ($p=3$) SD-LES method. Comparison with experimental measurements (PIV) \cite{bilka-PIV}. \label{fig:AvgW_Muffler}}
\end{figure}

\begin{figure}[t!]
\centering
    \subfigure[\small{$1 d$ downstream}.]{\includegraphics[width = 0.490\columnwidth]{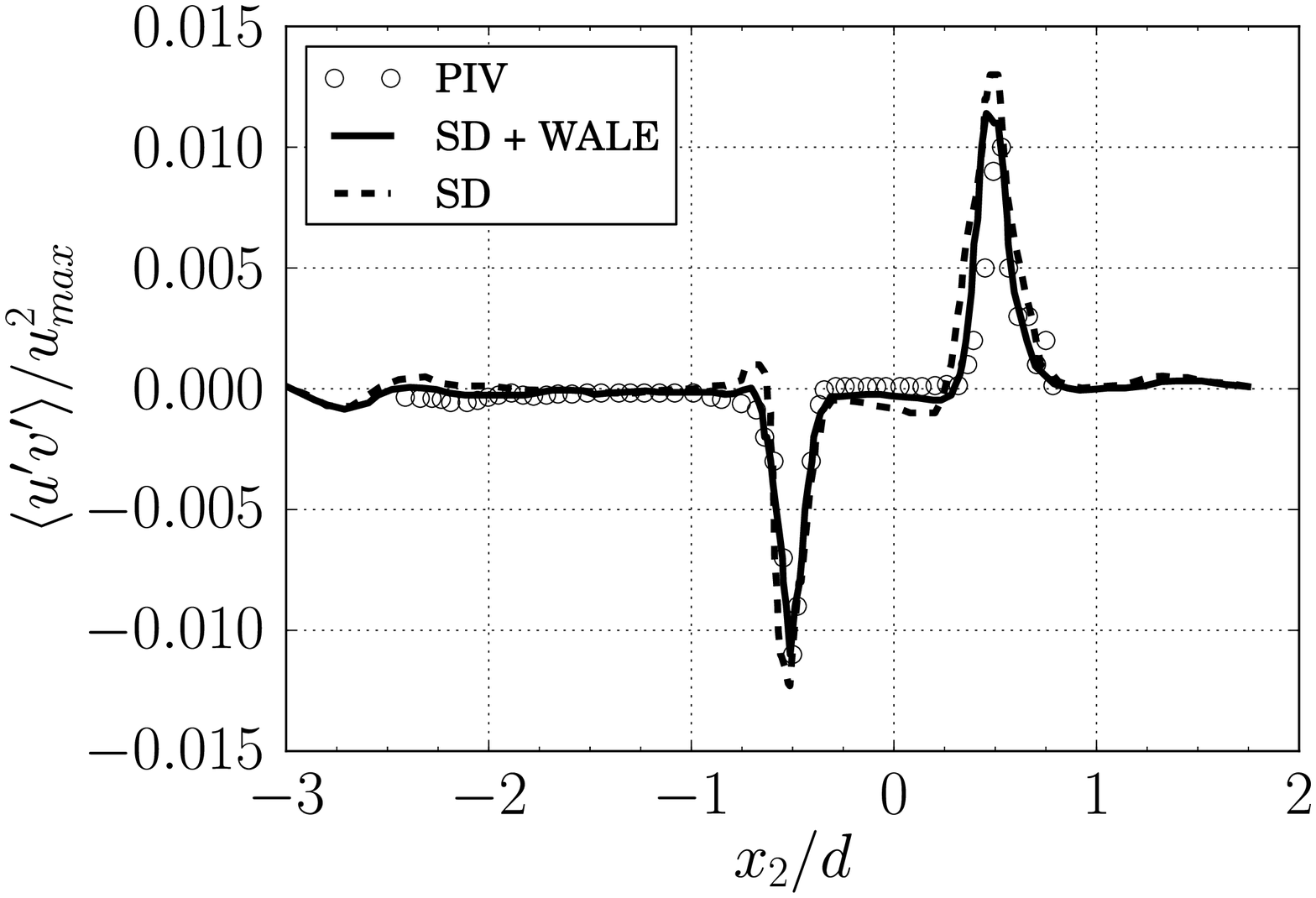}}  
	  \subfigure[\small{$4 d$ downstream}.]{\includegraphics[width = 0.490\columnwidth]{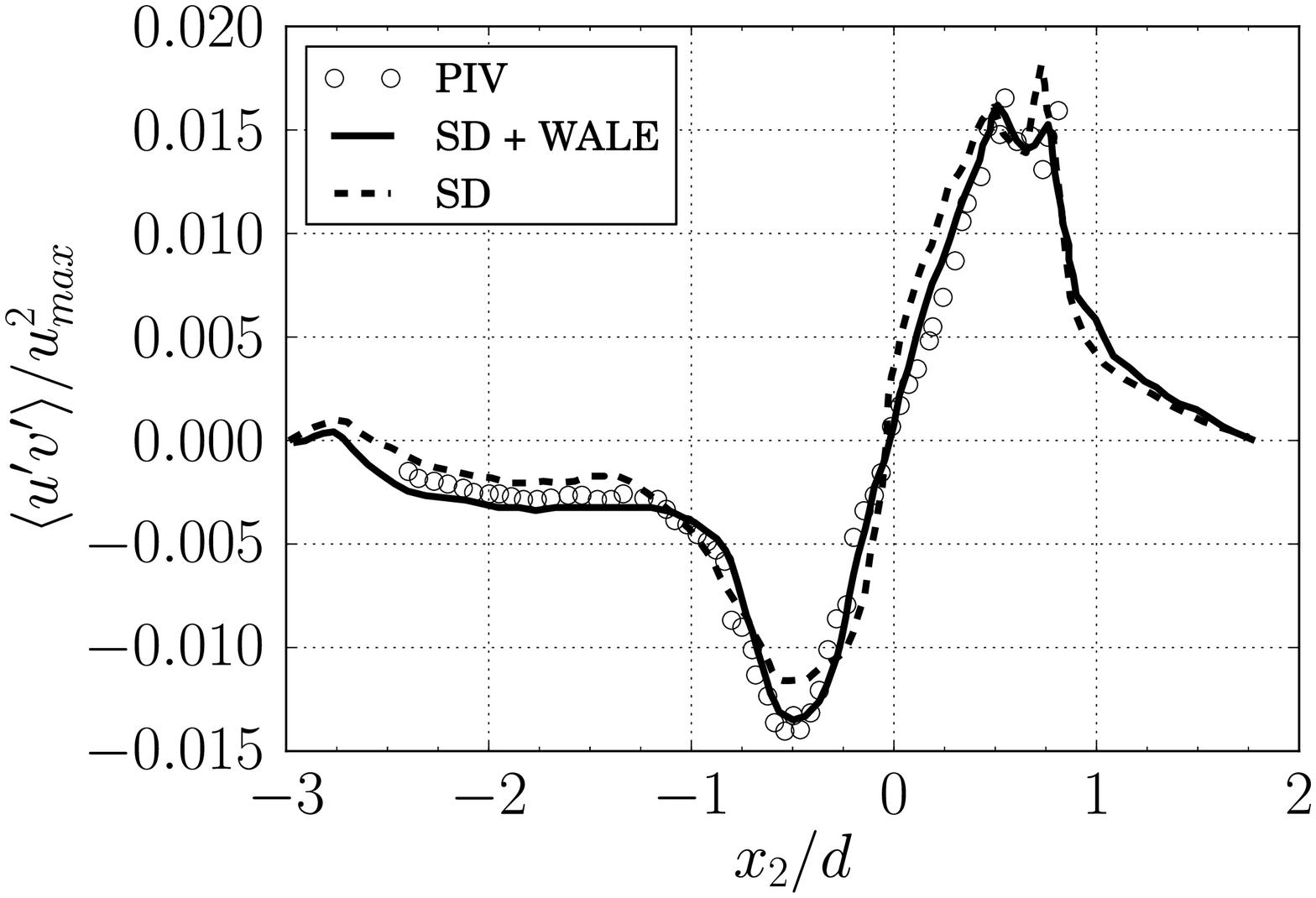}}
    \subfigure[\small{$6 d$ downstream}.]{\includegraphics[width = 0.490\columnwidth]{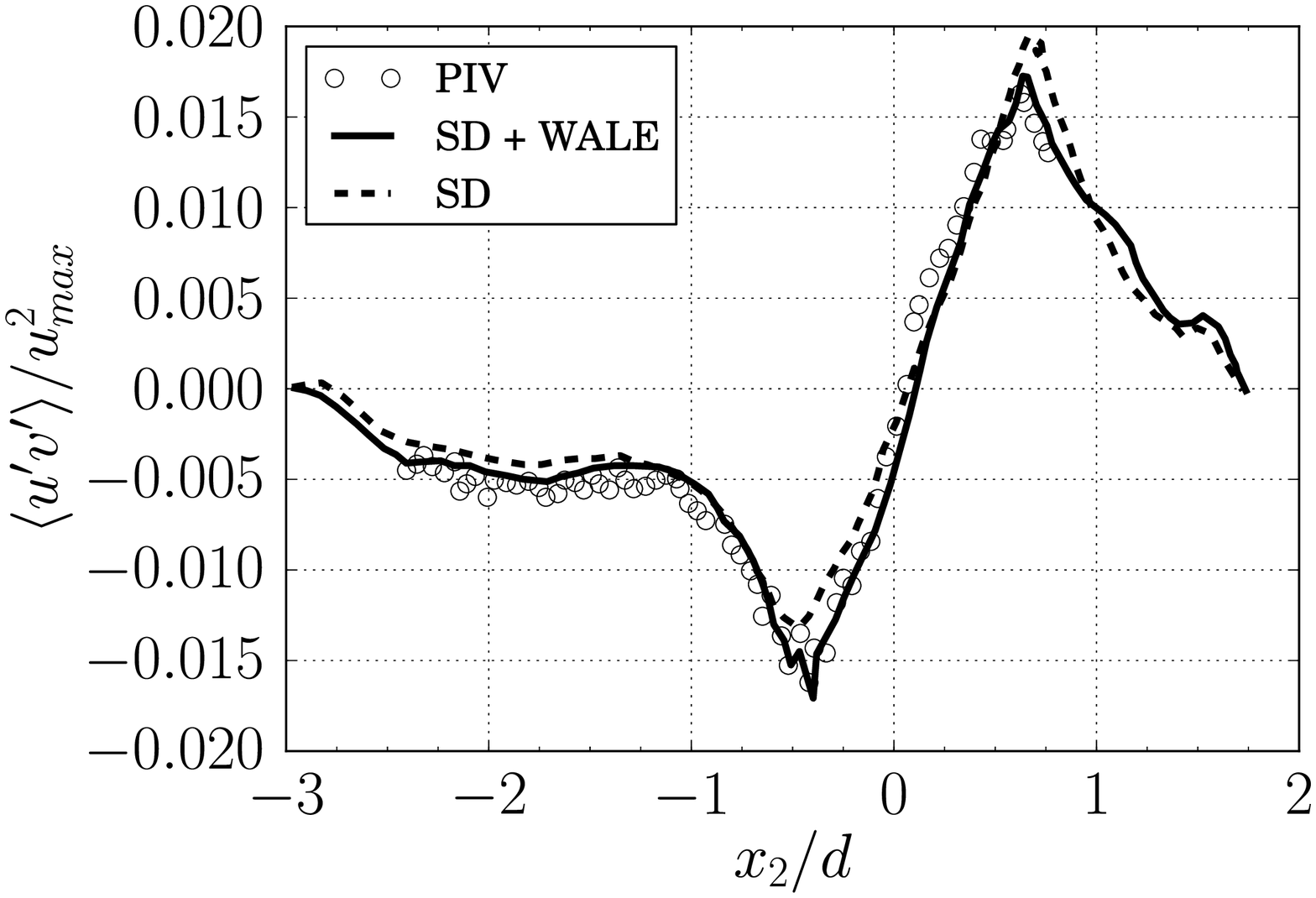}}  
    \subfigure[\small{$7 d$ downstream}.]{\includegraphics[width = 0.490\columnwidth]{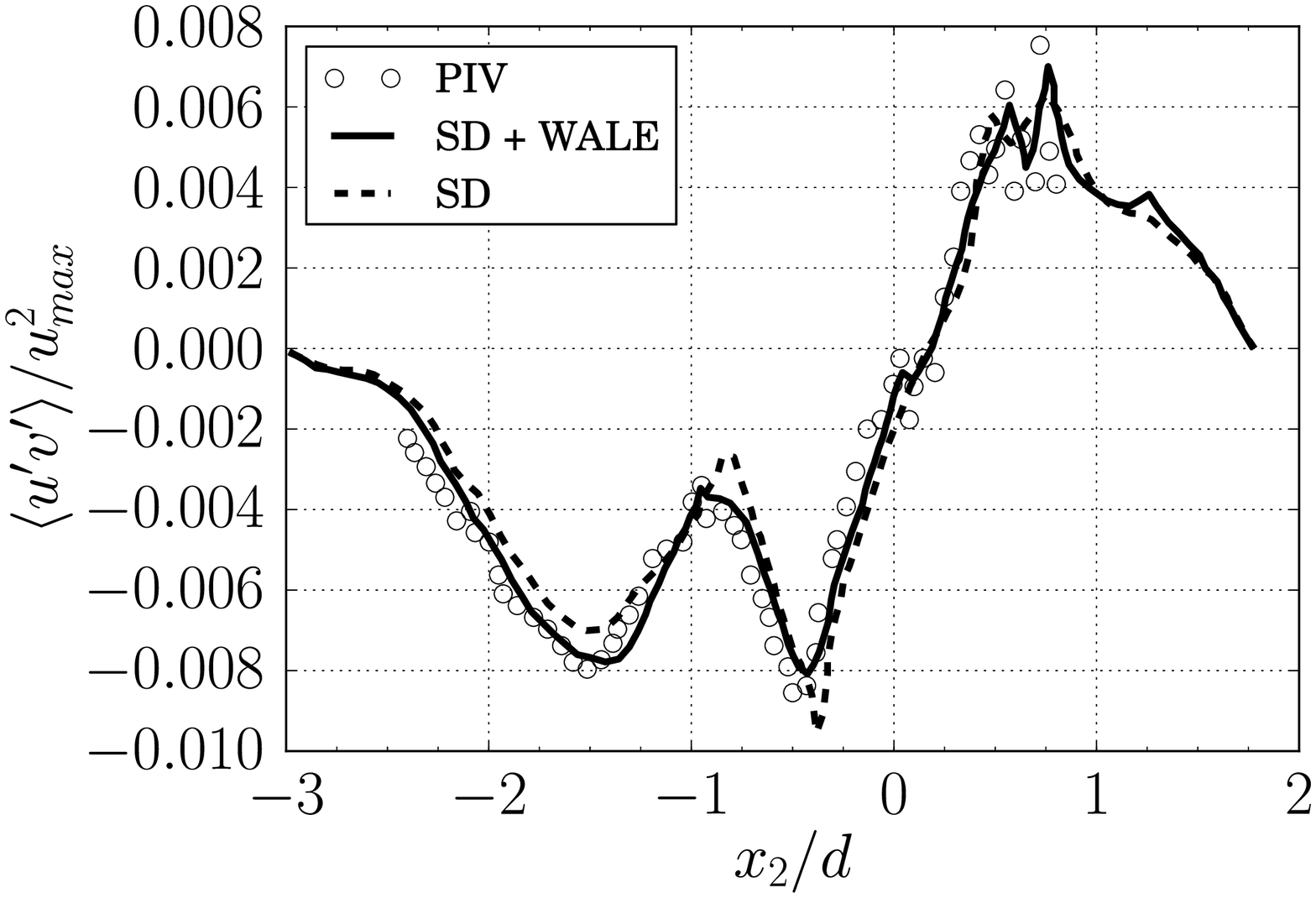}}
\caption{Reynolds stress $\langle u_2'u_3'\rangle/u_{max}^2$ in the axial direction at four cross sections in the expansion chamber, obtained with fourth-order ($p=3$) SD-LES method. Comparison with experimental measurements (PIV) \cite{bilka-PIV}. \label{fig:ReStrvw_Muffler}}
\end{figure}

In Figure \ref{fig:AvgW_Muffler}, the non-dimensional mean velocity profile in 
the axial direction $\langle\tilde{u}_3\rangle/u_{max}$ is shown for four 
different cross sections in the expansion chamber, where the PIV measurements 
were done. In this figure, the PIV data are also plotted for comparison. 
Figure \ref{fig:ReStrvw_Muffler} shows the non-dimensional Reynolds stress 
$\langle u_2' u_3' \rangle/{u_{max}^2}$ at the same cross sections. Although the
high-order implicit LES is already able to capture well the 
features of the flow field, the use of the WALE model improves 
the results. In particular, when the SGS model is active, the local extrema
of the time-averaged velocity profiles and the second-order statistical moment
(which get fairly oscillatory by moving far away from the inlet pipe) are better
captured.

\section{Conclusions}
\label{sec:conclusions}
The fourth-order SD method in combination with the WALE model  
and the variable filter width performed well. The numerical 
results confirm that the model is correctly accounting for the unresolved 
shear stress computed from the resolved field, for the present internal 
flow. However, it should be noted that the  
SD discretization without subgrid-scale modelling also worked rather well, at least for 
the grid resolution used in this study.

Work is currently under way to test both approaches for different orders of accuracy, grid resolutions
and other realistic turbulent flows. We 
believe that the flexibility 
of the high-order SD scheme on unstructured grids together with 
the development of robust sub-grid closure models for highly separated flows and efficient grid 
generators for high-order accurate 
schemes will allow to perform LES of industrial-type flows in the near future.

\section*{Acknowledgement}
The authors would like to thank Professor David I. Ketcheson and Professor Mark H. Carpenter
for their support. This research used partially the resources of the KAUST Supercomputing Laboratory and was supported in part by an appointment to the NASA Postdoctoral Program at Langley Research Center, administered by Oak Ridge Associates Universities. These supports are gratefully acknowledged.

%
%
%

\end{document}